# A New Route for the Determination of Protein Structure and Function

S. H. Mejias[1,2], A. L. Cortajarena[3,4], R. Mincigrucci[5], C. Svetina[1,6] and C. Masciovecchio[5]

[1]IMDEA-Nanoscience, Campus de Cantoblanco, 28049 Madrid, Spain

[2]Centro Nacional de Biotecnología (CNB-CSIC) – IMDEA Nanociencia Associated Unit "Unidad de Nanobiotechnología", Cantoblanco, 28049 Madrid, Spain

[3]Center for Cooperative Research in Biomaterials (CIC biomaGUNE), Basque Research and Technology Alliance (BRTA), Paseo de Miramón 194, Donostia-San Sebastián 20014, Spain.

[4]Ikerbasque, Basque Foundation for Science, 48009 Bilbao, Spain.

[5]Elettra Sincrotrone Trieste, S.S. 14 km 163, 34149 Basovizza, Trieste, Italy

[6]European XFEL, Holzkoppel 4, 22869 Schenefeld, Germany

## Abstract

Understanding complex biological macromolecules, especially proteins, is vital for grasping their diverse chemical functions with direct impact in biology and pharmacology. While techniques like X-ray crystallography and cryo-electron microscopy have been valuable, they face limitations such as radiation damage and difficulties in crystallizing certain proteins. X-ray free-electron lasers (XFELs) offer promising solutions with their ultrafast, high-intensity pulses, potentially enabling structural determination before radiation damage occurs. However, challenges like low signal-to-noise ratio persist, particularly for single protein molecules. To address this, we propose a new method involving engineered protein scaffolds to create ordered arrays of proteins with controlled orientations, aiming at enhancing the signal at the detector. This innovative strategy has the potential to address signal limitations and protein crystallization issues, opening avenues for determining protein structures under physiological conditions. Moreover, it holds promise for studying conformational changes resulting from photo-induced changes, protein-drug and/or protein-protein interactions. Indeed, the prediction of protein-protein interactions, fundamental to numerous biochemical and cellular processes, and the time-dependent conformational changes they undergo, continue to pose a considerable challenge in biology and biochemistry.

The determination of protein and macromolecular structures has traditionally relied on x-ray crystallography. This technique necessitates the growth of high-quality crystals, which must be both large enough to effectively diffract x-rays and withstand radiation damage (1). Indeed, protein crystallization is challenging due to the diverse properties of proteins, including poor solubility, conformational flexibility, and sensitivity to experimental conditions. Factors such as impurities, post-translational modifications and the need for trial-and-error optimization contribute to the difficulty. In the past few years, the technical developments of cryo-electron microscopy (cryo-EM) methodologies have triggered a resolution revolution in the single particle structural determination (2). Cryo-EM is a powerful imaging technique to study the detailed three-dimensional structure of biological macromolecules. This method involves imaging samples at extremely low temperatures to reduce radiation damage and preserve the structural integrity of the biological specimens. The process begins by flash-freezing the sample in a thin layer of vitreous ice, which helps to trap the molecules in their native, hydrated state. The electrons interact with the sample, and the resulting signals are collected to create 2D images. One of the primary limitations of cryo-EM lies in the resolution attainable for smaller proteins or complexes, where achieving atomic-level details becomes increasingly difficult. Another critical concern revolves around sample heterogeneity. Cryo-EM images represent an average of various conformations within a sample, and if there is significant variability in the structure or composition, it can compromise the overall resolution. This challenge becomes particularly pronounced when dealing with complexes that exhibit flexibility or multiple conformational states. In contemporary structural biology, crystallographic and cryo-EM data can complement each other in various ways, with two primary approaches being widely employed. Firstly, a low-resolution cryo-EM map of an entire molecule can provide the overall shape, and crystallographic atomic models of its components or homologs may be determined separately. These crystallographic models can then be docked into the cryo-EM map, allowing for the structure reconstruction of the entire molecule. On the other hand, a macromolecule can be crystallized and produce diffraction patterns with high resolution: the cryo-EM reconstruction of the macromolecule at a moderate resolution can be utilized as an initial model to solve the phasing problem of the crystals, enabling high-resolution structural determination. It is important to underline that both X-ray crystallography and cryo-EM are inherently unable to provide real-time depiction of dynamic processes, which prevents studying rapid conformational changes or ultrafast interactions. Moreover, the crystallization or the cryo treatment of protein brings the specimen into environments that may be far from the physiological one (3).

The recent development of artificial intelligence based platforms (AlphaFold) is allowing the determination of protein structures starting from their amino acid sequence and is tremendously helping in the prediction of protein structures (4). While AlphaFold represents a groundbreaking advancement, its current limitations include dependencies on existing structures, challenges with flexible regions, dynamic proteins and complexes, limited information on modifications, difficulties with membrane proteins, and constraints when dealing with large molecular assemblies. The most significant drawbacks of AlphaFold and similar machine-learning algorithms for structure prediction, stem from their reliance on learning patterns with limited knowledge of physics and chemistry. They can generate a single structure consistent with the learned patterns, but, currently, they lack the ability to provide a range of alternative conformations influenced by factors like pH, temperature or the binding of ions, ligands, or other proteins. Experiments are still essential for assessing these effects for which the proposed approach will allow to overcome current limitations (5).

Significant technical efforts pursued at X-ray Free Electron Lasers (XFELs) made it possible to recently carry out experiments of serial crystallography (SC) (6). XFELs are lasers that use a beam of accelerated electrons moving freely through a magnetic periodic structure to generate coherent and intense beams of light, spanning a wide range of wavelengths from the THz to the X-ray region. SC allows to obtain high resolution structural information of proteins by collecting the diffraction patterns of randomly oriented small protein crystals, usually impossible to characterize with synchrotron radiation. SC was initially envisioned as a stepping stone towards single particle imaging: one of the major science drivers for the creation of XFEs, based largely on the promise of coherent diffractive imaging methods and the concept of *diffract-before-destroy* (7). Atomic-scale structure determination of individual particles via coherent diffractive imaging (CDI) holds significant scientific promise for biology by revealing structure (and possibly dynamics) of complexes in near-physiological conditions. The concept proposed in (7) is currently hindered by technical limitations as, for instance, the very low number of photons diffracted and collected by the detector.

It is crucial to emphasize that none of the aforementioned methods enable the determination of a protein's structure under physiological conditions, particularly during interactions with other molecules such as drugs or other proteins. Of paramount importance is also understanding how external stimuli can affect the structure and dynamics of protein as, for example, the study of the dynamic behavior of photosystems for photoprotection (8). A deep understanding of these mechanisms is key for the photosystems behavior and their manipulation.

In this study, we propose a new sample delivery approach, based on the use of engineered scaffolds, involving the fabrication of sample supports designed to host the target protein in a near-native or dry environment. The concept revolves around achieving an array of aligned proteins enhancing the emitted signal well above the detectors sensitivity similarly to magnetic superlattices detected via soft X-ray diffraction or artificial nano-structures via hard X-ray diffraction. This would allow measurements of macrosystems in a hydrated environment maintaining conditions close to physiological settings during data acquisition similar to the idea previously proposed by our group (9). In that work, the proteins were presumed to be tethered to a substrate using techniques like split-intein mediated ligation. However, that method does not guarantee the complete absence of angular jitter among the deposited proteins. The ideal chemical binding method employed for protein attachment to the surface must ensure a 2D-ordered layer on the generated arrays, preserving the protein in its biologically active state. Nanopatterning and nano-deposition with nanoscale precision is nowadays possible with current technologies but, in general, requires expensive lithography methods and can only be done in relatively small sizes which makes this possibility quite impractical. On the other hand, the scaffolding approach introduced here is expected to reduce the conformational entropy of the arranged target proteins or complexes, thereby promoting conformational homogeneity in flexible proteins and complexes. By constraining degrees of freedom, scaffolded structures facilitate the study of dynamic proteins, allowing for easier characterization and analysis of structural dynamics. Moreover, this approach will also enable in situ studies of protein responses to binding events or interactions with drugs, providing valuable insights into biological processes at room temperature.

To generate the proposed arrays, we suggest a new methodology that uses a bottom-up approach combining site-specific protein modification and encoded protein self-assembly.

Based on this proposal, first, template protein scaffolds are modified with peptide binding sites to anchor the target protein of interest with site and orientation specificity (10,11). This anchoring strategy will generate linear nanostructures of the target protein that expand through different length scales (from ~ 3-20 nm at the molecular level to hundreds of nm (µm) at the supramolecular level) where the distance and orientation are controlled by the linking position engineered in the scaffold. For transferring the nanostructure to the macroscale, we will encode self-assembly properties to the template scaffolds, introducing interaction interfaces between them that generates ordered 2D or 3D-dimensional arrays. In these arrays, the nanostructure is expected to be ordered by peptide anchoring at the nanoscale. It has been demonstrated, through the assembly of efficient functional multi-enzymatic pathways, that this scaffolding strategy is compatible with the preservation of the structure and function of the scaffolded proteins (10).

Engineered proteins, and in particular Consensus Tetratricopeptide Repeat Proteins (CTPRs) (**Figure 1A**) are the perfect scaffold for the proposed approach. These proteins are composed of tandem arrays of 3 to 20 modular repeats that form superhelical structures of different lengths. The secondary and tertiary structure of these modular proteins are determined by a few amino acids. Thus, the rest of the amino acids can be modified without disrupting the protein structure. This structural control has made it possible to genetically encode protein functionalities to link nanoparticles or chromophores at specific protein positions, using engineered CTPR proteins as biomolecular templates (12–14). Moreover, these scaffolds show "head-to-tail" and "side-to-side" interactions that have been used to form nanostructured supramolecular materials (12–15). These inherent self-assembly properties have been engineered to encode tailored hierarchical assemblies such as linear fibers (16), 3D nanotubes (17), tightly packed monolayers (18), or anisotropic films (15,18). For the film formation, the most used fabrication strategy is by drop-casting, where a drop is deposited on a substrate dried under controlled conditions, allowing the proteins to self-assemble. In these systems the thickness of the film depends on the protein concentration and the drop size, however the films are not completely homogenous. A more accurate approach combining drop casting and spin coating allows to achieve a complete control over the thickness and homogeneity of the film, from monolayers of few nm-thickness to multilayers of µm-thickness (19). The methodology was optimized on silicon and $SiO_2$ substrates but can be translated to a substrate of choice upon optimization of depositing conditions, as we have demonstrated by depositing materials on gold substrate (12) and TEM carbon grids (unpublished data), among others.

For our purpose, we will take advantage of the functional flexibility and self-assembly of CTPR scaffolds to encode protein functional arrays with precise order (**Figure 1**). A CTPR16, with 16 TPR repeats, will be decorated with Tetratricopeptide Affinity Repeat Proteins (TRAPs), making it selective template scaffolds for the peptide-tagged target proteins (10,11) (**Figure 1A**). The target protein is linked to the CTPR16 scaffold selectively by following a key lock strategy (**Figure 1B**). The size of the CTPR16 allows us to control the distance and orientation of the target protein at the nanoscale to optimize the nanostructure having at least 7.5 nm between proteins on an initial design leaving enough space for protein conformational freedom (**Figure 1C**). We will use the self-assembly properties of the CTPR to scale the order from the nanoscale to the macroscale (**Figure 1C**). The proposed approach allows us to generate 2D and potentially 3D arrays of well-ordered proteins for the X-ray experiments. While 2D arrays are attractive for studying protein structure and dynamics as all proteins face an external layer that is free to interact, 3D structures can be attractive to further enhance the X-ray signal following the same interference process as in Bragg diffraction generated in bulk crystals. Finally, CTPR scaffolds could be additionally engineered to modify their assembly properties by engineering protein-protein contacts, achieving scaffolds that assemble selectively in one or two dimensions, which could be useful to decrease the conformational entropy of flexible proteins.

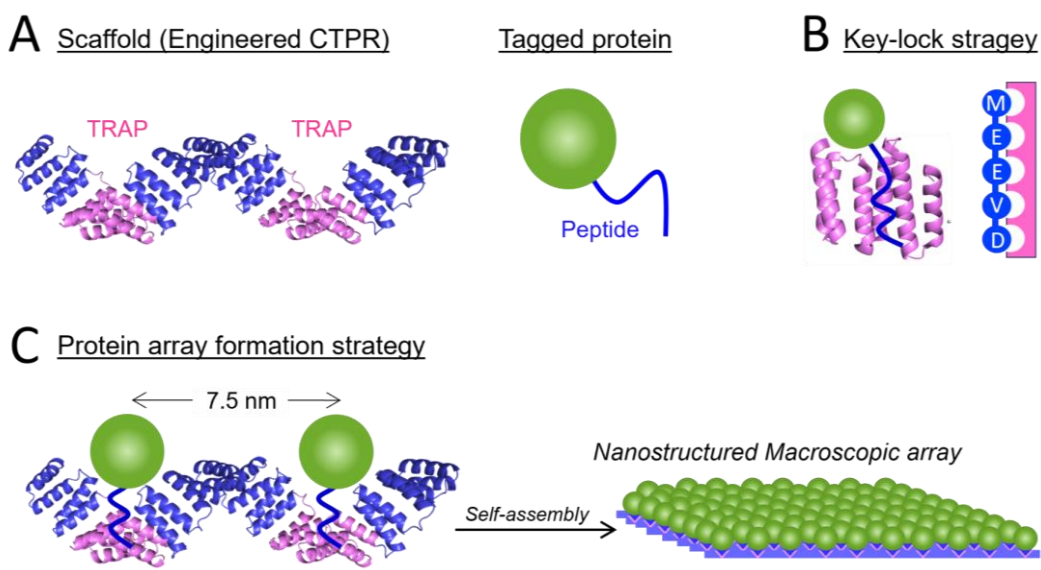

**Figure 1. A.** Designed CTPR scaffold with TRAPs peptide binding sites in pink. Tagged targeted protein with peptide in blue. B. TRAP scaffold-based key-lock strategy with the engineered MEEVD peptide sequence for TRAPs binding. C. Targeted protein nanostructure ordered by engineered CTPR protein. Macroscopic array formation leaded by CTPR protein self-assembly.

We propose to validate the potential of our method of aligned protein samples using two complementary diffraction techniques that have recently been employed to study 2D crystals: 1) Grazing Incidence X-ray Diffraction (GIXD, Figure 2A) and XFEL radiation (Figure 2B).

GIXD has been developed at synchrotrons (21). This technique is used for surface-sensitive studies of thin films and surfaces, providing detailed structural information. Its shallow penetration depth enhances signal-to-noise ratios, making it valuable for analyzing crystalline structures, epitaxial growth, and dynamic processes in materials. GIXD has been already used to measure 2D arrays of membrane proteins obtained by adsorbing proteins to ligand-lipid monolayers at the surface of water (22). However, the latter process can be applied only to a few membrane proteins and while strategies have been proposed to ameliorate it (22), the achieved resolution stops at 10 Å due to the dynamic disorder. This is the case when using cross-linkers that reduce the thermally activated structure fluctuations that are the main responsible of the dynamic disorder.

XFEL radiation is a complementary approach to GIXD, that also add high temporal resolution. It is demonstrated that XFEL radiation is able to identify diffraction peaks from bacteriorhodopsin two-dimensional crystals mounted on a solid support and kept at room temperature (23). While this technique permits the use of standard Bragg diffraction setups, the high pulse intensities required to generate signals in the detector will inevitably destroy the sample upon irradiation following the concept of diffract-before-destroy (7). This is not the case for GIXD, where longer acquisition times are feasible without detrimental effects on the proteins. Bacteriorhodopsin is unique in that it is naturally produced in two-dimensional crystalline patches (24). The generalization of this kind of measurement to artificially ordered 2D protein assembly is conceptually straightforward and points toward the feasibility of our approach. The resolution in (23), limited to 7 Å, was primarily constrained by the small signal generated from the 2D submicrometric bacteriorhodopsin crystals, resulting in insignificant intensities at the large-angle Bragg peaks. However, with our proposed approach, 2D crystals of about 1 cm$^2$ area can be easily produced thus permitting a raster scan of the sample significantly boosting the data statistics. For instance, when considering samples of 100 square millimeters, this could potentially yield a factor of, at least, $10^3$ increase in the integrated signal with respect to what obtained in (23) where 11 images out of 334 were retained acceptable. Another important advantage would be the use of attosecond pulses that are recently produced by XFELs (25). Indeed, it has been demonstrated that after about 500 attoseconds, sample radiation damage occurs during diffraction (26) signifying that photons impinging on the sample after this time interval will not produce the desired signal and, eventually, decrease the signal-to-noise ratio. Taking into account the experimental signal reported in (23), where the X-ray pulse duration was approximately 50 femtoseconds, we can estimate a reduction in the scattered signal of at least a factor of $10^2$. Thanks to recent developments in the physics and technology of XFELs (27,28) it has been shown that pulses with time duration of 200 attoseconds and 200 µJ energy are easily produced at 9 keV (J. Yan et al., in preparation). Considering that in (23) the pulse intensity was of 2 mJ, this implies an increase of another factor 10 in the number of photons collected by the detector. The use of beam spot sizes larger than 100 µm could also sensibly reduce the radiation damage of the sample/substrate. In fact, considering that in (23) the 2 mJ XFEL pulses were focused to an area of 0.1 µm$^2$, the interaction between the very high electric field strengths and atoms in the protein causes plasma formation in less than 1 fs (29).

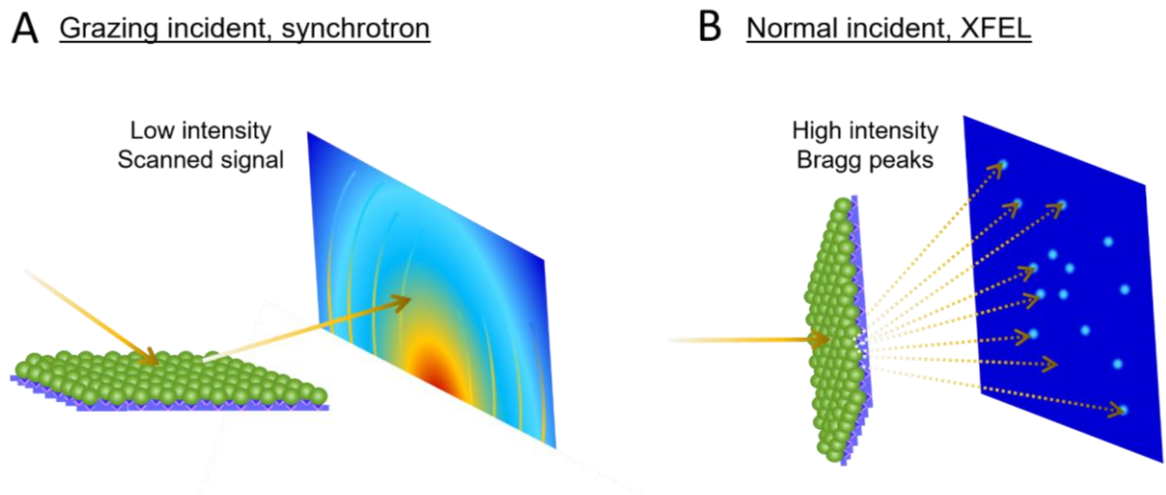

**Figure 2 . A)** X-ray diffraction experiment based on grazing incidence experiments using synchrotron radiation. **B)** Bragg peaks obtained in XFEL experiments based on a high-energy single shot.

In these proposed methodologies, both the scaffold and the protein/complex of interest contribute to the generation of Bragg peaks. While this may introduce some complexity in the structural determination of the specimens, the ability to measure the signals separately from the scaffold layer could aid in distinguishing signals from the two systems. Furthermore, it could serve as a reference field to assist in solving the phase problem during reconstruction similarly to holography where a known spherical wave is artificially generated and used as a reference (30). The substrates that can be used may range from vitreous silica, which produce diffuse scattering or crystalline silicon wafers generating Bragg peaks that can be set out of the interesting diffraction region or, eventually, used as a reference.

To further evaluate the feasibility of our idea, we present a case study focused on a specific target protein. We consider the Dronpa protein and its variants ideal candidates for examining our strategy. Dronpa is a fluorescent protein that can be switched reversibly between a fluorescent (on-state) and a non-fluorescent (off-state) state, by illuminating it with light of varying wavelengths in the optical range. This light-induced structural transformation involves both trans-to-cis isomerization and proton transfer of its chromophore. Dronpa has been tailored through rational engineering to enhance its switching properties, making it valuable in various applications such as super-resolution microscopy (31–33), nanoscopy (34), and optogenetics (33). However, despite these advancements, the precise mechanisms, time scales and relative contributions of chromophore and protein dynamics remain poorly understood, posing challenges for further protein engineering efforts (35). Our approach can potentially elucidate the molecular basis of the switching mechanism. Indeed, Dronpa's characteristics make it an ideal scaffold for integrating with our system for two reasons: 1) its rigid β-barrel structure limits conformational flexibility which helps for a proper structure determination by enhanced protein signal in scaffold arrays; 2) light can be used as an external stimulus to trigger chromophore isomerization with the corresponding protein conformational changes that can be tracked by optical pump - X-ray probe. Overall, our approach involves two main steps: firstly, employing molecular biology techniques to introduce suitable peptide modifications for scaffold linking with Dronpa, and secondly, utilizing steady-state X-ray diffraction to determine its structure and pump-probe X-ray methods to elucidate its switching cycle. By pursuing this methodology, we aim to shed light on the molecular underpinnings of the Dronpa switching mechanism, key knowledge that will allow to optimize its structure for applications by rational design. In a broader context, our methodology exhibits significant potential to measure the conformational changes of natural and synthetic enzymes under physiological conditions showing the key steps of their function which now can be only addressed based on theoretical calculations (36,37). Furthermore, our strategy holds promise for shedding light on the photoactivation mechanisms underlying the dynamic behavior of de novo designed photosystems —an essential aspect of their development, albeit challenging due to the small size and inherent flexibility of these proteins (38). In this sense, our proposed scaffolding strategy presents a compelling avenue for enhancing the conformational homogeneity of flexible proteins and achieving a unified conformation across multiple conformers. By carefully designing the target protein peptide labeling site, linker length, and the linking position on the CTPR scaffold, we can favor a specific protein conformation while preserving the protein´s inherent conformational freedom. This reduction in conformational entropy facilitates the streamlined exploration of protein dynamics in response to external stimuli such as drug binding or light exposure, providing a notable advantage over existing strategies. This distinct advantage underscores the potential of our method to advance comprehension in the realm of protein dynamics and potentially catalyze the development of novel therapeutic interventions. Given these considerations, the strategy outlined in this paper holds the potential to address key challenges in protein structure determination and dynamic behavior with impact in a broad range of research fields, including nanotechnology, photosynthesis, or biomedicine.

In this article, we present an innovative methodology for delivering two-dimensional, aligned arrays of biomolecules. Our method extends the applicability of structural biology techniques to macromolecules, especially proteins, that are traditionally challenging to crystallize. It potentially enables the utilization of synchrotron and X-ray Free-Electron Laser to reconstruct protein structure in physiological conditions and study its function while interacting with ligands in real-time from femtosecond up to longer times. Experiments can be conducted under near-native conditions, providing insights into biological processes in environments that mimic real-life scenarios and allowing the study of protein-protein as well as protein-drug interactions in real-time. The approach facilitates pump-and-probe experiments, enabling optical laser excitation of the protein of interest and subsequent XFEL beam probing for structural changes post-excitation. Finally, our approach requires only a moderate sample quantity, making it advantageous for challenging or expensive-to-produce proteins. Overall, this proposed methodology represents a comprehensive and multidisciplinary strategy with broad implications for advancing structural biology techniques, particularly in understanding the structures of challenging macromolecules under physiological conditions and during dynamic processes.

## Acknowledgments

S.H.M. acknowledge the funding from the Spanish Ministry of Science and Innovation (project TED2021-131906A-I00), and her fellowship from "La Caixa" Foundation (ID 100010434). The project is also supported by a 2023 Leonardo Grant for Researchers and Cultural Creators, BBVA Foundation (LEO23-2-9635). The BBVA Foundation accepts no responsibility for the opinions, statements and contents included in the project and/or the results thereof, which are entirely the responsibility of the authors. IMDEA Nanociencia acknowledges support from the 'Severo Ochoa' Programme for Centres of Excellence in R&D of the Spanish Ministry of Science and Innovation (CEX2020-001039-S).